\documentclass{article}

\usepackage{PRIMEarxiv}

\usepackage[utf8]{inputenc} 
\usepackage[T1]{fontenc}    
\usepackage{hyperref}       
\usepackage{url}            
\usepackage{booktabs}       
\usepackage{amsfonts}       
\usepackage{nicefrac}       
\usepackage{microtype}      
\usepackage{lipsum}
\usepackage{fancyhdr}       
\usepackage{graphicx}       
\graphicspath{{media/}}     

\usepackage{amssymb}
\usepackage{mathtools}
\usepackage{amsmath}
\usepackage{graphicx}
\usepackage{float}
\usepackage{epsfig}
\usepackage{csquotes}
\usepackage[font=normalsize,labelfont=normalsize]{caption}
\usepackage{subcaption}
\usepackage{cleveref}
\usepackage[T1]{fontenc}
\usepackage{color}
\usepackage{lettrine}
\usepackage{url}
\usepackage{makecell}
\usepackage{enumitem}
\usepackage[section]{placeins}
\usepackage[table]{xcolor}  
\usepackage{array}
\usepackage{easybmat}
\usepackage{multirow,bigdelim}
\usepackage{diagbox}
\usepackage{multicol}

\usepackage{float}
\usepackage{algorithm,algorithmic}
\usepackage{nicefrac}

\usepackage{setspace}
\usepackage{tensor}
\usepackage{appendix}

\usepackage{tikz}
\usepackage[utf8]{inputenc}
\usepackage{pgfplots} 
\usepackage{pgfgantt}
\usepackage{pdflscape}
\pgfplotsset{compat=newest} 
\pgfplotsset{plot coordinates/math parser=false}

\newlength\fwidth
\newlength\fheight

\usepackage{amsthm}

\title{Explainable AI using\\Inherently Interpretable Components\\for Wearable-based Health Monitoring}

\author{
  Maurice Kuschel\textsuperscript{1} \\ \texttt{maurice.kuschel@sst.uni-paderborn.de} \And
  Solveig Vieluf \textsuperscript{2, 3} \\ \texttt{solveig.vieluf@med.uni-muenchen.de} \And
  Claus Reinsberger \textsuperscript{4, 5} \\ \texttt{reinsberger@sportmed.uni-paderborn.de} \And
  Tobias Loddenkemper \textsuperscript{6} \\ \texttt{tobias.loddenkemper@childrens.harvard.edu} \And
  Tanuj Hasija \textsuperscript{1} \\ \texttt{tanuj.hasija@sst.uni-paderborn.de} \\
  \\ \\
  \textsuperscript{1}Signal and System Theory Group, Paderborn University, Paderborn, Germany \\
  \textsuperscript{2}Department of Medicine I, Ludwig Maximilian University, Munich, Germany \\
  \textsuperscript{3}Konrad Zuse School of Excellence in Reliable AI, Munich, Germany \\
  \textsuperscript{4}Institute of Sports Medicine, Paderborn University, Paderborn, Germany \\
  \textsuperscript{5}Division of Sports Neurology and Neurosciences, Mass General Brigham, Boston, USA \\
  \textsuperscript{6}Division of Epilepsy and Clinical Neurophysiology, Boston Children’s Hospital,\\ Harvard Medical School, Boston, USA
}

\newcommand{\model}{M}
\newcommand{\timeseries}{\mathbf{x}}
\newcommand{\representation}{\mathbf{r}_{\mathbf{x}}}
\newcommand{\components}{{\mathbf{C}_{\mathbf{x}}}}
\newcommand{\decompf}{F}
\newcommand{\decompfinv}{F^{-1}}
\newcommand{\weights}{\mathbf{w}}
\newcommand{\baseline}{\mathbf{b}}

\begin{document}

\maketitle

\begin{abstract}
The use of wearables in medicine and wellness, enabled by AI-based models, offers tremendous potential for real-time monitoring and interpretable event detection. Explainable AI (XAI) is required to assess what models have learned and build trust in model outputs, for patients, healthcare professionals, model developers, and domain experts alike. Explaining AI decisions made on time-series data recorded by wearables is especially challenging due to the data's complex nature and temporal dependencies. Too often, explainability using interpretable features leads to performance loss. We propose a novel XAI method that combines explanation spaces and concept-based explanations to explain AI predictions on time-series data. By using Inherently Interpretable Components (IICs), which encapsulate domain-specific, interpretable concepts within a custom explanation space, we preserve the performance of models trained on time series while achieving the interpretability of concept-based explanations based on extracted features. Furthermore, we define a domain-specific set of IICs for wearable-based health monitoring and demonstrate their usability in real applications, including state assessment and epileptic seizure detection.
\end{abstract}

\section{Introduction}

Health monitoring with wearable devices, such as smartwatches or fitness bands, is becoming increasingly popular \cite{piwek2016rise}. Continuous, real-time monitoring of the autonomic nervous system (ANS) enables proactive, preventive health management. In addition to personal applications, such as fitness, activity, or sleep tracking, an increasing number of medical applications are emerging as medical-grade wearables develop. These devices enable non-stigmatizing methods of monitoring medical conditions. In patients with neurological disorders such as epilepsy or Parkinson's disease, wearables can already be used to detect seizures and track tremors \cite{yu2023artificial, Poleur2025Wearable}. Cardiac patients can use continuous heart-rate monitoring to detect atrial fibrillation\cite{Barrera2025Accuracy} and to track blood pressure\cite{mukkamala2022cuffless} via wearables. Although these state-of-the-art methods deliver promising performance, they are often based on deep neural networks that are not interpretable \cite{ribeiro2016should}. While this problem is not specific to these applications, it is particularly pronounced in this area, as it can lead to severe consequences such as overdose, loss of trust among clinicians and patients, and affect the quality of life of patients \cite{amann2020explainability}.

The field of eXplainable AI (XAI) focuses on opening this black box and making neural network predictions interpretable \cite{ali2023explainable}. However, interpreting AI decisions on time-series data is a non-trivial problem \cite{rojat2021explainable}. Compared with other data types such as images or tables, time series are inherently complex, and existing methods — particularly saliency-based explanation techniques — struggle to extract intuitive explanations \cite{schroder2023post}. Typically, saliency-based methods work by perturbing the input data, for example, in the time \cite{pmlr-v139-crabbe21a, pmlr-v202-enguehard23a, liu2024explaining} or frequency domain \cite{vielhaben2024explainable, brusch2025freqrise}. While these approaches do not require retraining the model and operate on time series data, thereby making no performance trade-offs, they answer the question "Where is the important region?" under the assumption of localized, sparse information. However, in time series, these explanations often raise more questions than they answer, e.g., why specific parts of the data are essential and whether the results are important in isolation or in context with other data. Especially when multiple regions or modalities are highlighted simultaneously, the exact combination of those and the temporal dependence of the time points remain unknown. In addition, generalizing multiple local regions of interest to global explanations is not trivial, either.

Concept-based XAI methods specifically address this problem. Given a set of concepts, these methods generate explanations by assigning importance values to each concept. Furthermore, multiple instance-level explanations can be easily amalgamated. While their explanations are more intuitive, the underlying concepts must still be provided, often during training \cite{koh2020concept}. Further, relying solely on concepts can reduce model performance because it does not fully exploit the temporal structure of time-series data. Alternatively, analyzing a trained model using representative example sets is possible as well \cite{kim2018interpretability}. While this approach allows for post-hoc explanations of concepts, its practical implementation is not straightforward. Defining high-quality example sets that well represent a concept is not trivial. Overall, concept-based methods provide more intuitive, even global explanations, but require manual intervention and often compromise model accuracy. Ideally, we aim to combine complementary methodological strengths: the simplicity and performance of saliency methods with the interpretability of concept-based explanations.

Recent approaches already aim to bridge the gap between saliency- and concept-based explanations by leveraging explanation spaces \cite{rezaei2024explanation, brusch2024flextime}. These methods acknowledge that selecting a representation space aligned with human intuition is essential for producing helpful explanations. They propose using invertible transformations on such spaces to make explanations more accessible to domain experts. However, despite operating in these spaces, they still apply saliency-based methods, thereby failing to deliver actual concept-based explanations.

To address this issue, we propose a novel, model-agnostic method using Inherently Interpretable Components (IICs). Drawing on the concept of reversible explanation spaces, we propose creating decomposition spaces in which each component is naturally interpretable to domain experts, and no further local XAI method is required. By focusing on the importance attribution of components rather than timestamps or frequency bands, the explanations become sparser and more straightforward, provided the decomposition space is interpretable. Furthermore, those explanations can be readily aggregated into global explanations, thereby enabling cohort-level analyses. 

To apply our proposed explanation technique to health monitoring with wearables, we also propose a set of IICs for monitoring the ANS, with components explicitly selected for each modality recorded by a wearable. We selected concepts from the literature on ANS health monitoring and implemented the corresponding decompositions to align with the method's constraints. We demonstrate our proposed method on two real-world wearable datasets, including state assessment and epileptic seizure detection, trained on different network architectures.

Our contributions in this paper are as follows:
\begin{enumerate}[label=\arabic*)]
    \item We propose a novel, concept-based, model-agnostic XAI method that utilizes IICs to generate local (instance-level) and global (cohort-level) explanations for models predicting on raw time series.
    \item We propose a decomposition of multimodal wearable data into IICs, enabling application of the proposed method to ANS data.
    \item We apply the method to generate local and global explanations for time-series neural networks trained on real-world wearable data, including participant state assessment and epileptic seizure detection, using Long Short-Term Memory (LSTM) and transformer models, respectively.
\end{enumerate}

\section{Related Work}

\subsection{Saliency-Based Explanations}

Saliency methods are widely used to identify informative regions in the input by assigning feature-wise importance scores. While these methods were developed initially for tabular data, they have recently been adopted for other domains, such as time series \cite{pmlr-v139-crabbe21a, pmlr-v202-enguehard23a, liu2024explaining}. It has been shown that the same approach of assigning importance scores to individual timestamps does not yield the same level of interpretability due to the more complex nature of time-series data and temporal dependencies between timestamps \cite{rojat2021explainable}. Explanations that assign importance to temporal positions can fail when class-distinctive features are not temporal but latent parameters. In those cases, saliency scores did not reveal the latent features that actually drive the predictions \cite{schroder2023latent}. 

\subsection{Explanation Spaces}

Explanation spaces are designed to overcome this problem \cite{schroder2023latent, rezaei2024explanation, brusch2024flextime, vielhaben2024explainable}. \cite{rezaei2024explanation} proposes the general use of other spaces that provide more interpretable explanations by capturing the latent features relevant to the prediction. As long as the transformation is reversible, it can offer an explanation space. The original time series is projected into an explanation space, where existing XAI techniques can be used to generate explanations on relevant components. Due to invertibility, the data can be projected back to the input space, so the neural network does not need to be retrained. This enables highly flexible, model-agnostic post-hoc explainability. Furthermore, no trade-off in model performance is required, as the model is trained on raw time-series data.

Several studies have already focused on inspecting the frequency and time-frequency domains \cite{brusch2024flextime, vielhaben2024explainable}. Virtual inspection layers transform time series to interpretable representations via short-time Fourier transformation \cite{vielhaben2024explainable} and propagate relevance scores through layer-wise relevance propagation. FLEXTime \cite{brusch2024flextime} uses a bank of filters to decompose the original signal into interpretable frequency bands and apply saliency-based XAI methods to them. While these approaches address the problem that class-distinctive features may not be temporal parameters \cite{schroder2023latent}, they work only when an interpretable set of frequency bands exists. Therefore, this applies only to a small subset of latent features. Another major drawback is that it still requires an additional XAI method to be applied in the explanation space. Therefore, the core problem that positional information may not capture the actual features still exists, just in a different space.

\subsection{Concept-Based Explanations}

Concept-based explanations use high-level, human-understandable concepts rather than low-level input features \cite{koh2020concept, kim2018interpretability}. This results in highly intuitive, domain-expert-aligned explanations. Concept-based explanations can be more easily amalgamated into global, cohort-level explanations, as a concept is defined across multiple samples.

One way to apply XAI concepts is to do so a priori, before model training. Approaches like Concept Bottleneck Models (CBMs) \cite{koh2020concept} use constrained models that learn concepts from raw data as an intermediate step, then make predictions based on them. By imposing this constraint, the method established an explicit link between learned concepts and predictions, thereby making it inherently interpretable. While its interpretability is appealing, this method requires a dataset with additional concept labels for training, and does not allow changing the concepts post-hoc. Secondly, while the bottleneck is introduced for interpretability reasons, it can also impact the performance. The model is no longer end-to-end and cannot use the full temporal information in the time series, as it routes all information through the concept bottleneck. Another way to utilize concepts in explanations is to analyze models in a post-hoc fashion. Approaches such as testing with concept activation vectors (TCAV) \cite{kim2018interpretability} provide concept-based explanations without requiring retraining. They define concepts as sets of representative input examples and analyze the model's activations when the selected inputs are fed through it. While this approach requires fewer concept labels, its implementation is comparatively complex, as selecting inputs for each concept must be carefully done to represent the concept of interest accurately. Furthermore, the method assumes that concepts are linearly separable in the model's latent space, which is not guaranteed, especially for complex or entangled concepts.

\section{Proposed XAI Method}
The goal is to develop an explanation method that combines complementary methodological strengths: the simplicity and performance-preserving properties of saliency-based methods, along with the intuitive interpretability of concept-based methods. To achieve this, we propose decomposing the time series into interpretable components, so that no additional local XAI techniques are required; a simple importance assignment suffices. Instead of focusing on specific timestamps or regions, we emphasize the importance of whole components, which domain experts can assess intuitively. We can also combine components into global rules, as they encapsulate generalizable concepts. This way, we get global, concept-based explanations at the cohort level without performance degradation or retraining.

\begin{figure}
    \centering
    \includegraphics[width=0.6\linewidth]{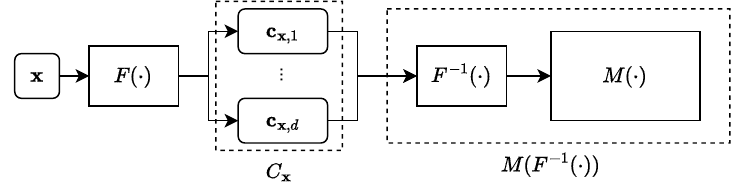}
    \caption{Visualization of decomposition and the extended network pipeline. The original time series $\timeseries$ gets decomposed into components $\components$ via $\decompf$. The inverse transformation $\decompfinv(\cdot)$, combined with the original time series network $\model$, results in a new, extended model $\hat{\model}$ that uses interpretable components as inputs, without retraining.}
    \label{fig:decomp_pipeline}
\end{figure}

Let $ \timeseries = \{ x_1, ..., x_t \} \in \mathbb{R}^{t}$ be a time series of length $t$. Assume we have a trained black-box model $\model$, which maps the time series $\timeseries$ to another representation $\representation$, with $\representation \in \mathbb{R}^s$. We decompose our time series into a set of components $\components = \decompf(\timeseries) = \{\mathbf{c}_{\timeseries,1},...,\mathbf{c}_{\timeseries,d}\}$, where $d$ is the number of components. To create concept-based explanations in an explanation space while still utilizing the original model $M$ trained on raw time-series data, we require our decomposition function $\decompf(\cdot)$ to be invertible with $\decompfinv(\cdot)$, such that $\decompfinv(\components) = \timeseries$. As proposed in \cite{rezaei2024explanation}, we can define an extended model, $\hat{\model} = \model(\decompfinv(\components))$, that takes $\components$ as an input, utilizes $\decompfinv(\cdot)$ to reconstruct a time series and feed it to $\model$. This enables feeding components of interest into the original model, without requiring retraining. See Figure \ref{fig:decomp_pipeline} for visualization.

As this approach enables us to use signal components as inputs to the model trained on time-series data, we can also generate explanations in this component space. In contrast to \cite{rezaei2024explanation}, we propose not to use local XAI methods on the signal components. By assigning importance scores to each component, we can provide concept-based explanations that highlight its significance and relevance. To achieve this, we attempt to remove as many components from the original signal as possible without significantly altering the neural network's output. To give explanations concerning the components in $\components$, we introduce a weight vector $\weights \in \mathbb{R}^{d}$, with $\weights = [w_1, ..., w_d]^T$ and $0 \leq w_i \leq 1$. Then, $\decompfinv(\components \weights)$ creates a weighted reconstruction $\timeseries'$ from $\timeseries$. With a weight vector of only ones, $\decompfinv(\components \weights)$ reconstructs the original input $\timeseries$. We propose an explanation based on weighted composition, similar to that in \cite{brusch2024flextime}, albeit not exclusively in the frequency domain but in general explanation spaces.

To eliminate components from the explanation space, we introduce a dedicated weight loss term, $L_{weights}$. It is computed by taking the sum of all weights in the weight vector $\weights$, normalized by the total number of weights, $d$:
\begin{align}
    L_{weights} = \frac{1}{d} \sum_{i=1}^d \weights_i.
\end{align}
While minimizing the weights in the vector $\weights$ eliminates components, we only want to remove negligible ones, those that are not relevant to the network output. To achieve this, we introduce an additional loss term, $L_{degradation}$, for output degradation. We compute it by comparing the model output for the altered data, $\model(\decompfinv(\components \weights))$, to the original output, $\model(\timeseries)$. The loss is computed via the absolute difference of the outputs:
\begin{align}
    \text{degradation} = | \model(\decompfinv(\components \weights)) - \model(\timeseries) |.
\end{align}
While this evaluates the degradation numerically, we also aim to specify an upper bound on the output degradation. We are interested in component importance only when we assume the network will continue to predict in a manner similar to the original network. Therefore, we apply an upper bound to the degradation by adding a penalty, $p$, to the degradation that exceeds the allowed maximum, $\text{max}_{deg}$: 
\begin{align}
    L_{degradation} = p \cdot \max (0, \text{degradation} - \text{max}_{deg}).
\end{align}
If the degradation is higher than $\text{max}_{deg}$, the penalty $p$ is applied. In any other case, $L_{degradation}$ is zero. The combination of both loss terms allows us to eliminate all components from the signal that are not relevant for the model's prediction:
\begin{align}
    L_{IIC} = L_{weights} + L_{degradation}.
\end{align}

When eliminating components in the explanation space, we must consider one additional factor. A general problem when perturbing input data for a trained neural network is the lack of conformity with the training data. When removing all components, we may produce a time series that influences the network output only because it does not match the input distribution on which the network was trained. For instance, assume a network is trained on data from a sensor that reports patients' heart rates. When all components are removed, the resulting signal is an array of zeros. While this is technically correct, the network was trained on a specific distribution of heart-rate inputs, and we must ensure that the reconstructed data remains within an acceptable domain when modified. To account for this and decouple the components of interest from the conformity component, we propose using a baseline component, $\baseline$, based on the mean of the training data in each decomposition to obtain a valid signal. This baseline component is excluded from the optimization function. This is similar to how other XAI methods use calibration data to establish a baseline for the dataset \cite{lundberg2017unified}. As this baseline is always required, it does not have a weight that can be adjusted in the optimization.

The resulting explanation, $\weights_{\timeseries}$, is a sparse weight vector with weights in $[0,1]$, indicating which components are negligible, $w_i = 0$, and which ones are not, $w_i > 0$, for a specific time series $\timeseries$. To facilitate post-processing of these explanations, such as creating group-level explanations or identifying frequent itemsets from multiple inputs, we can apply thresholding to $\weights_{\timeseries}$ to obtain a binary importance vector. The corresponding pseudocode for the complete optimization is presented in Algorithm \ref{alg:optimization}.

\begin{algorithm}
\caption{Optimization}\label{alg:optimization}
{\setlength{\parskip}{0pt}
\begin{algorithmic}[1]
    \STATE \textbf{Input:} $\model$, $\timeseries$, $\decompf$, $\decompfinv$, $p$, $\text{max}_{deg}$
    \STATE $\components \gets \decompf(\timeseries)$;
    \STATE Initialize $\weights_{\timeseries} \gets \mathbf{1}$;
    \WHILE{stopping criterion is not reached}
        \STATE $\text{outp}_{orig} \gets \model(\timeseries)$;
        \STATE $\text{outp}_{alt} \gets \model(\decompfinv(\components \weights_{\timeseries}))$;
        \STATE $\text{degradation} \gets | \text{outp}_{orig} - \text{outp}_{alt} |$;
        \STATE $L_{degradation} \gets p \cdot \max (0, \text{degradation} - \text{max}_{deg})$;
        \STATE $L_{weights} \gets \text{mean}(\weights_{\timeseries})$;
        \STATE $L_{IIC} \gets L_{weights} + L_{degradation}$;
        \STATE $\weights_{\timeseries} \gets \text{backprop}(\weights_{\timeseries}, \nabla_{\weights_{\timeseries}} L_{IIC})$;
    \ENDWHILE
\end{algorithmic}}
\end{algorithm}

To analyze the computational complexity of Algorithm \ref{alg:optimization}, assume we explain $n$ samples of $\timeseries$ and run the while loop for $e$ epochs. The most relevant operations to focus on are the forward paths of the trained model, $\model(\timeseries)$, and the inverse decomposition, $\decompfinv(\components \weights_{\timeseries})$. As their computational complexity depends on the actual model architecture and choice of components, we denote them as $\mathcal{O} (\model(\timeseries))$ and $\mathcal{O} (\decompfinv(\components \weights_{\timeseries}))$. The backward passes are denoted by $\nabla \model(\timeseries)$ and $\nabla \decompfinv(\components \weights_{\timeseries})$ and bear the cost of computing gradients via backpropagation. By the standard analysis of reverse-mode automatic differentiation, $\mathcal{O} (\nabla \model(\timeseries)) = \mathcal{O} (\model(\timeseries))$ and $\mathcal{O} (\nabla \decompfinv(\components \weights_{\timeseries})) = \mathcal{O} (\decompfinv(\components \weights_{\timeseries}))$. That is, gradient computation has the same asymptotic complexity as the forward pass \cite{goodfellow2016deep}. Therefore, the resulting computational complexity is given as $\mathcal{O} ( e \cdot n \cdot ( \model(\timeseries) + \decompfinv(\components \weights_{\timeseries}) ) ) $. Complexity for a specific case can then be computed by considering $\mathcal{O} (\model(\timeseries))$ for the respective model architecture \cite{goodfellow2016deep, hochreiter1997long,vaswani2017attention}, $\mathcal{O}( \decompfinv(\components \weights_{\timeseries}) )$ for the implementation of the reconstruction function, the number of timestamps $t$, and the number of components $d$.

\section{Inherently Interpretable Components for the Autonomic Nervous System (ANS)}
\label{sec:iics}

In this section, we propose a transformation, $F_{ANS}(\cdot)$, to extract IICs from wearable recordings that monitor the ANS for health monitoring. We consider data from wearable devices, such as smartwatches, equipped with built-in sensors that record multiple modalities, including accelerometry (ACC), electrodermal activity (EDA), wrist or ankle skin temperature (TEMP), and heart rate (HR).

In the following sections, we will consider each sensor individually. We will discuss the sensor's purpose, what it measures, and how these measurements are used in the literature to identify patterns. We will then focus on deriving IICs for a concept-based explanation space.

\begin{figure}
    \centering
    \includegraphics[width=0.6\textwidth]{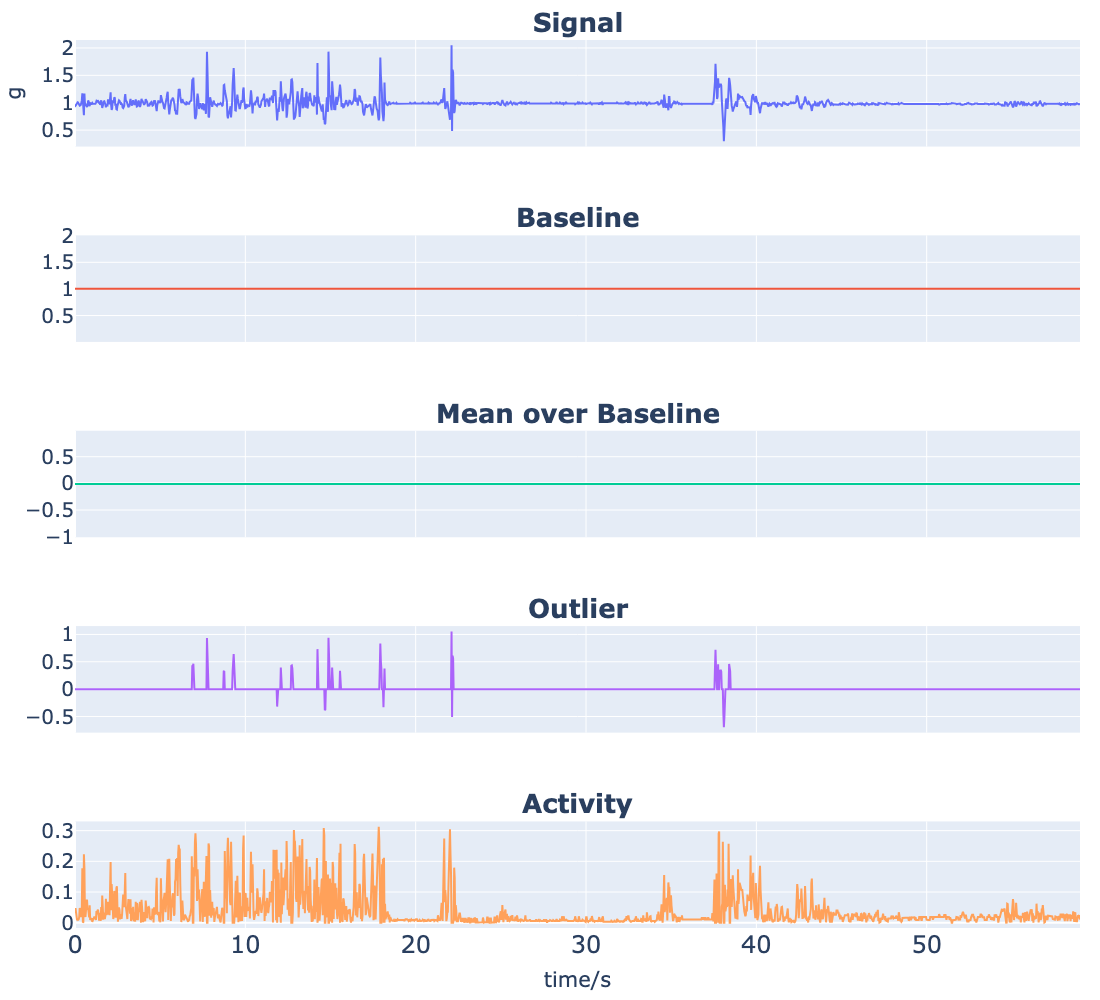}
    \caption{Exemplary decomposition of an accelerometer signal into IICs.}
    \label{fig:iic_acc}
\end{figure}

\textbf{Accelerometer} data is recorded to recognize movement and activity patterns. Accelerometer data are typically recorded along three axes (X, Y, and Z), which correspond to left-right, forward-backward, and up-down motion, respectively. While these sensors do not directly measure movement or velocity, they measure the total acceleration forces acting on the sensors. This includes gravitational acceleration, a static component, and motion-based acceleration, a dynamic component, measured in $g$. In the literature, various methods have been proposed to extract activity indicators from ACC data \cite{Osipov2015Objective, Ksiazek2024Assessment, Costantini2024Artificial}. The overall approach is usually to first summarize the data from all axes into a unimodal version called the resultant acceleration. Let $r_i = \sqrt{x_i^2 + y_i^2 + z_i^2}$ be the resultant acceleration for timestamp $i$, with $x_i$,$y_i$, and $z_i$ being the values from the X, Y, and Z axes at timestamp $i$. Then, this resultant acceleration will be split into the static, gravitational component and the dynamic, activity component. The latter component can then be processed using different statistical features to enable motion recognition. \cite{vaha2015universal} compared different metrics based on the difference to the mean level and found the Mean Amplitude Deviation (MAD) over a time window to be the most robust. It describes the typical distance of data points from the mean. Given a resultant ACC timeseries $\mathbf{r} = [r_0, ..., r_t]$ of length $t$, the MAD can be computed as $\text{MAD} = \frac{1}{t} \sum_{i=0}^t| r_i - \bar{r} |,$ where $\bar{r}$ is the mean value of the time series. For our concrete transformation of ACC data into interpretable components, we propose following the pattern in the literature by first decomposing the resultant acceleration into a static and a motion-based component. The static, gravitational component is computed as $\bar{r}$, the mean value of $\mathbf{r}$. To account for conformity with the seen training data, we further separate it into the baseline, $\baseline_{acc}$, which is the mean of the training data, and the additional mean-over-baseline component, $\mathbf{c}_{meanob}$. For motion-based acceleration, we first apply z-scoring \cite{grubbs1969procedures} to separate outlier movement, $\mathbf{c}_{outlier}$. For the activity component, we adopt the MAD \cite{vaha2015universal} to capture movement. However, as we require our decomposition to be invertible, we cannot simply extract mean values over windows from a timeseries, because we would not be able to revert it to a timeseries. Therefore, we generate an approximate time-series signal by excluding window averaging and using the absolute deviation from the mean at each timestamp, yielding $\mathbf{c}_{activity}$. We preserve the sign of the deviation during the optimization procedure for reconstructing the original input in the inverse. Pseudocode for the decomposition is presented in Algorithm \ref{alg:acc}. An exemplary decomposition of an ACC signal is shown in Figure \ref{fig:iic_acc}.

\begin{algorithm}
\caption{$F_{acc}(\mathbf{r}_{acc}, \baseline_{acc})$}\label{alg:acc}
\begin{algorithmic}[1]
    \STATE $\mathbf{c}_{meanob} \gets \text{mean}(\mathbf{r}_{acc} - \baseline_{acc})$;
    \STATE $\mathbf{r}_{zero-mean} \gets \mathbf{r}_{acc} - \baseline_{acc} - \mathbf{c}_{meanob}$;
    \STATE $\mathbf{c}_{outlier} \gets \text{z-score}(\mathbf{r}_{zero-mean})$;
    \STATE $\mathbf{c}_{activity} \gets | \mathbf{r}_{zero-mean} - \mathbf{c}_{outlier} |$;
\end{algorithmic}
\end{algorithm}

\textbf{Heart rate} data is recorded for a wide range of health applications, often using optical sensors using photoplethysmography (PPG). By measuring the reflection of an LED light, which is affected by the blood pulse through the arteries under the skin, the sensor returns the PPG signal. This signal can be used to extract multiple signals, such as RR intervals and heart rate, expressed in beats per minute. In the literature, several features are used to describe cardiac activity. They can be divided into two main categories: overall heart rate level \cite{Vieluf2020Autonomic, Osipov2015Objective} and heart rate variability metrics, such as root mean square of successive differences (RMSSD), and low-/high-frequency components \cite{Singh2022Observational, Ksiazek2024Assessment, Soares2020Parkinsons, Wilson2025Abnormal}. 

\begin{figure}
    \centering
    \includegraphics[width=0.6\textwidth]{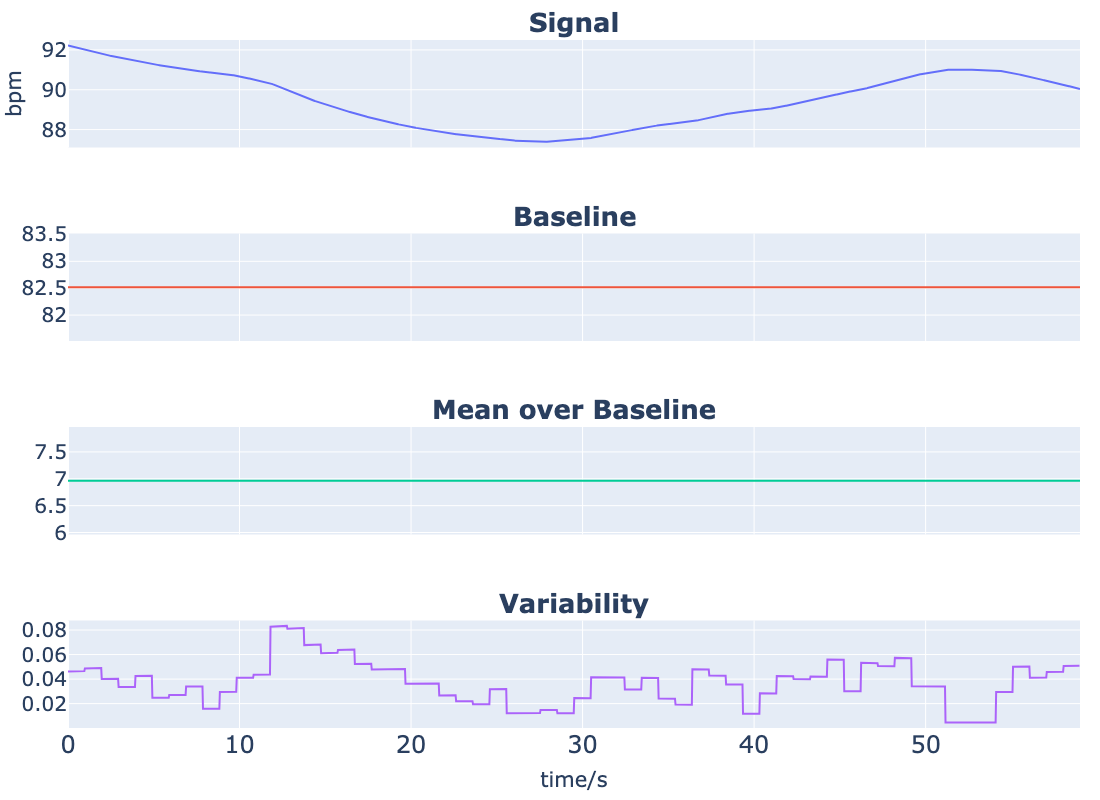}
    \caption{Exemplary decomposition of a heart rate signal into IICs.}
    \label{fig:iic_hr}
\end{figure}

We propose a decomposition of the heart rate data as follows: heart rate variability features, such as SDNN and RMSSD, are concerned with the length of RR intervals. Therefore, we first transform our heart rate data from the BPM domain to the RR domain by applying $T_{RR}(\mathbf{x}) = \frac{60.000}{\timeseries}$ to the heart rate values, as we want to convert the data from $bpm$ to intervals, measured in $ms$. The result is a time series containing RR interval lengths in milliseconds, $\timeseries_{RR}$. To capture the overall RR interval level, we create a mean component derived from averaging the entire series. Then, we transform it back to the BPM domain and subtract an HR baseline component, $\baseline_{hr}$, to ensure validity, yielding the mean-over-baseline component, $\mathbf{c}_{meanob}$. To encapsulate the short-term changes in the RR data, we would compute the RMSSD. However, since we require an invertible transformation, summarizing a whole time series in a single value is not possible. Therefore, we define our HRV component, $\mathbf{c}_{variability}$, as the difference between consecutive RR intervals, thereby encapsulating RMSSD information in a time series. A visualization of an exemplary decomposition is shown in Figure \ref{fig:iic_hr}.

\begin{algorithm}
\caption{$F_{hr}(\timeseries_{hr}, \baseline_{hr})$}\label{alg:hr}
\begin{algorithmic}[1]
    \STATE $\timeseries_{RR} \gets T_{RR}(\timeseries_{hr})$;
    \STATE $\mathbf{c}_{meanob} \gets T_{RR}(\text{mean}(\timeseries_{RR})) - \baseline_{hr}$;
    \STATE $\mathbf{c}_{variability} \gets | \text{diff}(\timeseries_{RR}) |$;
\end{algorithmic}
\end{algorithm}

\textbf{Electrodermal activity}, also known as galvanic skin response (GSR), measures changes in the skin's conductance caused by sweat activity underneath the wearable device. The sensor typically measures conductance in units of microSiemens ($\mu S$). Sweat activity is controlled by the sympathetic nervous system and can therefore be used to estimate the sympathetic arousal \cite{dawson2007electrodermal}. In the literature, various features are computed from raw data to assess arousal. While time features are already insightful \cite{Vieluf2020Autonomic, Touserkani2018T147, Zsom2019Ictal, Kim2018Automatic, Sempeles2012Wrist}, a commonly used scheme is to decompose the signal into tonic and phasic components by considering components from different frequency bands \cite{Ramirez2021Ictal, Andrikopoulos2024Machine, Costantini2024Artificial}. Tonic EDA, also known as skin conductance level, captures slow baseline shifts. Phasic EDA, also known as skin conductance response, captures rapid, event-related spikes in skin conductance. Because the two components measure complementary information, analyzing them individually is more insightful.

\begin{figure}
    \centering
    \includegraphics[width=0.6\textwidth]{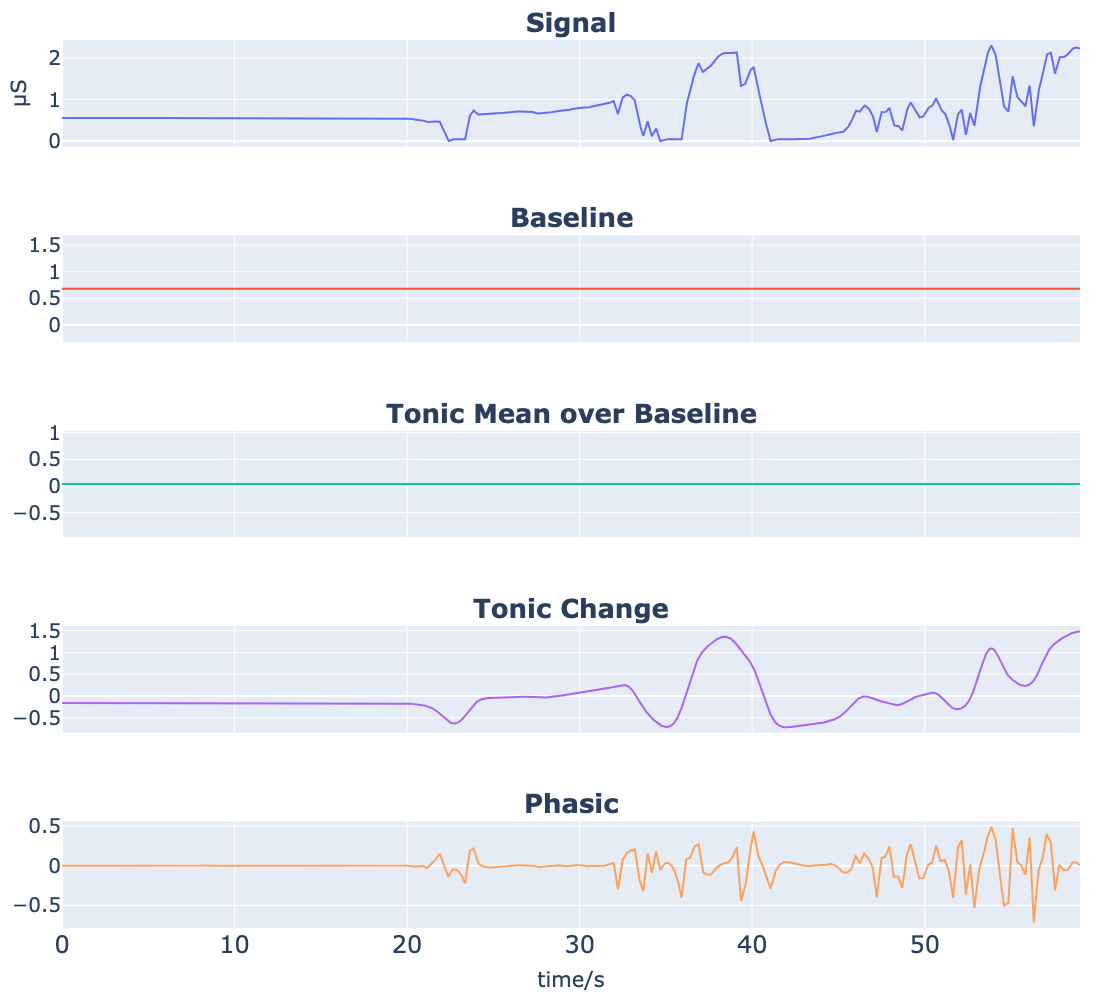}
    \caption{Exemplary decomposition of an EDA signal into IICs.}
    \label{fig:iic_eda}
\end{figure}

For generating IICs, we follow this approach and use the \textit{NeuroKit} package \cite{Makowski2021neurokit}, written in Python, to decompose the EDA signal into tonic and phasic components. However, because the tonic signal captures the overall level of electrodermal activity, including the baseline $\baseline_{eda}$, and exhibits slow changes, we apply an additional decomposition to separate these components. Overall, we end up with the following components: tonic mean over baseline, $\mathbf{c}_{tonic,meanob}$, tonic change, $\mathbf{c}_{tonic, change}$, and phasic, $\mathbf{c}_{phasic}$. Algorithm \ref{alg:eda} contains the pseudocode for the decomposition. An exemplary decomposition of an EDA signal can be found in Figure \ref{fig:iic_eda}.

\begin{algorithm}
\caption{$F_{eda}(\timeseries_{eda}, \baseline_{eda})$}\label{alg:eda}
\begin{algorithmic}[1]
    \STATE $\timeseries_{tonic} \gets \text{NeuroKit.tonic}(\timeseries_{eda})$;
    \STATE $\mathbf{c}_{tonic,meanob} \gets \text{mean}(\timeseries_{tonic} - \baseline_{eda})$;
    \STATE $\mathbf{c}_{tonic,change} \gets \timeseries_{tonic} - \baseline_{eda} - \mathbf{c}_{tonic,meanob} $;
    \STATE $\mathbf{c}_{phasic} \gets \text{NeuroKit.phasic}(\timeseries_{eda})$;
\end{algorithmic}
\end{algorithm}

\textbf{Skin temperature} data measures the temperature of the skin under the wearable. Although it is not identical to core body temperature, skin temperature still provides valuable insights into a patient's physiological state and circadian rhythm, reflecting changes in health, stress, and sleep \cite{charkoudian2003skin}. Skin temperature sensors often measure infrared radiation emitted by the skin and convert it into a temperature reading, reported in degrees Celsius or degrees Fahrenheit. Components of interest reported in the literature include signal statistics, such as mean and variance, as well as entropy \cite{Vieluf2020Autonomic, Pepa2024Supervised}.

\begin{figure}
    \centering
    \includegraphics[width=0.6\textwidth]{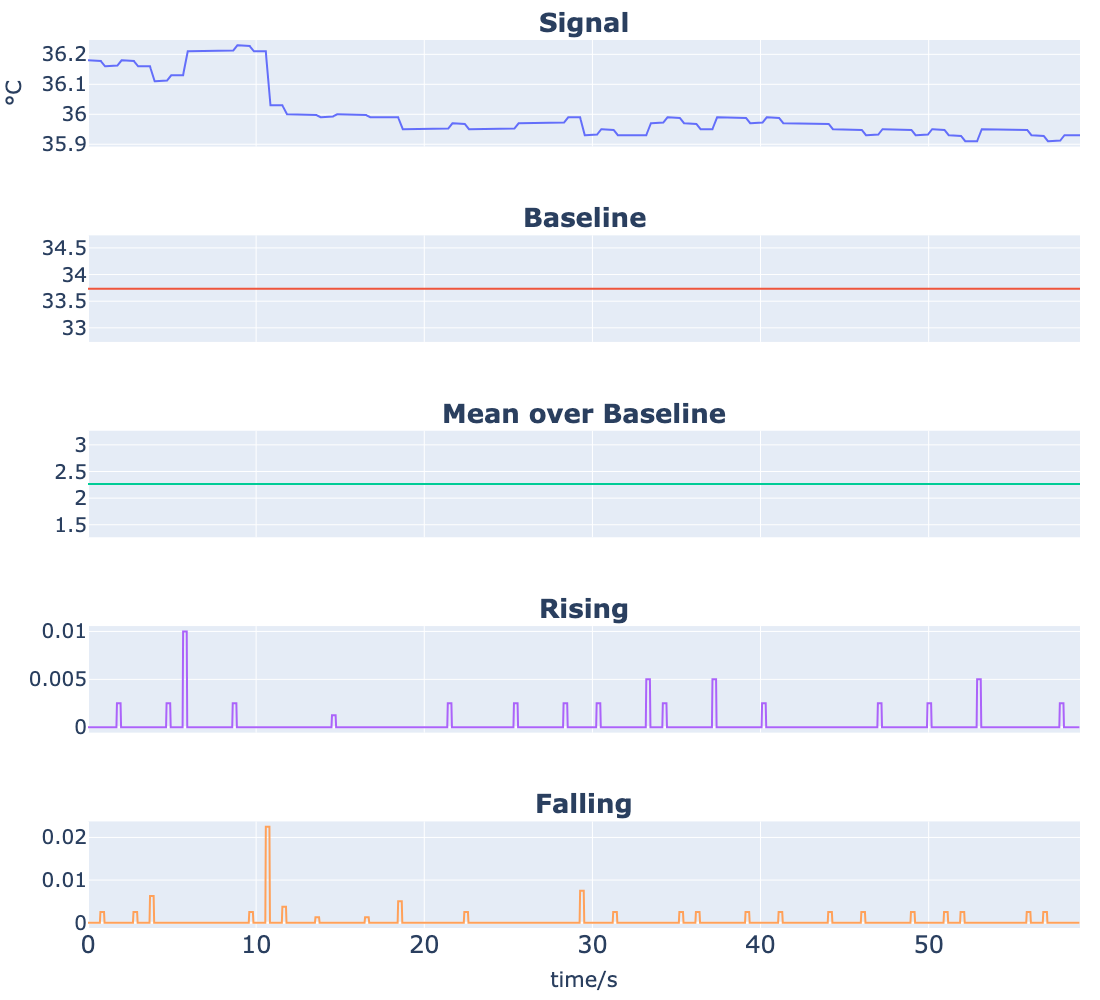}
    \caption{Exemplary decomposition of a temperature signal into IICS.}
    \label{fig:iic_temp}
\end{figure}

We propose decomposing the temperature data into static and dynamic components, following a procedure similar to that used for other modalities. As no temperature-specific metric has been proposed in the literature, we extract the static component by computing the mean relative to the baseline, $\mathbf{c}_{meanob}$, which characterizes the overall level in relation to the temperature baseline of the training set, $\baseline_{temp}$. Given that multiple references suggest that temperature shifts are relevant as event-driven responses \cite{charkoudian2003skin}, we further introduce a dynamic temperature component. We capture temperature changes by computing differences between consecutive temperature values and splitting them into two components based on their sign. This yields two components that capture the rising and the falling part, $\mathbf{c}_{rising}$ and $\mathbf{c}_{falling}$, respectively. An exemplary decomposition is shown in Figure \ref{fig:iic_temp}.

\begin{algorithm}
\caption{$F_{temp}(\timeseries_{temp}, \baseline_{temp})$}\label{alg:temp}
\begin{algorithmic}[1]
    \STATE $\mathbf{c}_{meanob} \gets \text{mean}(\timeseries_{temp} - \baseline_{temp})$;
    \STATE $\text{diffs} \gets \text{diff}(\timeseries_{temp} - \baseline_{temp} - \mathbf{c}_{meanob})$;
    \STATE $\mathbf{c}_{rising} \gets [d \text{ if } d > 0 \text{ else } 0 \text{ for } d \in \text{diffs}]$;
    \STATE $\mathbf{c}_{falling} \gets [d \text{ if } d < 0 \text{ else } 0 \text{ for } d \in \text{diffs}]$;
\end{algorithmic}
\end{algorithm}

\section{Experiments}

To assess the performance of the proposed explanation method, we apply it to two real-world wearable datasets and compare it with baseline approaches using multiple metrics to evaluate numerical performance and explanatory power.

\subsection{Datasets}

\subsubsection{State Assessment Through Wearables}
In our first experiment, we will apply all approaches to data from a wearable study conducted with informed consent by the Sports Medicine Group at Paderborn University (Validierung von Handgelenkssensoren zur Erfassung der Aktivität im Autonomen Nervensystem, Nr. 56/2024). Participants were equipped with an Empatica E4 wearable. They underwent two phases during which ANS data, as described in Section \ref{sec:iics}, were recorded: rest in the supine position and stepping. In our experiments, we used this data to create a relatively simple classification problem, for which the indicators distinguishing the classes are well established in the literature. Given the simplicity of the classification problem, we consider only a subset of modalities, namely heart rate, EDA, and temperature, since accelerometry alone can perfectly distinguish the classes \cite{yang2010review}. 

We used data from 14 participants, with 7.5 minutes per participant and class. The raw time-series signals were windowed into 30-second non-overlapping segments. We use LSTM models for our IIC approach, which were trained in a $7$-fold setting with 2 participants per fold. We used the evaluation dataset to stop training and load the model with the best performance. For each fold, we use the best model from five different initializations. We report the performance and explanation metrics on the evaluation dataset, averaged over all folds. 

\subsubsection{Epileptic Seizure Detection With Wearables}
In our second experiment, we apply the approaches to the task of epileptic seizure detection with wearable data. We use data from patients enrolled at the Boston Children's Hospital (BCH) (Autonomic biomarkers of seizures to assess the risk for SUDEP, IRB-P00001526), with informed consent, including data from patients enrolled in the long-term monitoring unit with video-EEG who wore an Empatica E4 sensor on their wrist/ankle
From the entire patient group, we included patients with a confirmed epilepsy diagnosis and those who had tonic-clonic seizures. The data were preprocessed, filtered for data quality, and segmented into ictal (i.e., during a seizure), preictal, postictal, and interictal (i.e., without seizure activity) regions. Further details on preprocessing, data quality assessment, and segmentation are provided in \cite{hasija2025improving}. In our experiments, the filtered dataset comprises wearable data from 37 patients and 61 seizures.

The seizure detection task involves distinguishing between ictal and interictal time-series windows to determine whether a patient is currently experiencing a seizure. We create a balanced dataset by pairing each ictal segment with an interictal segment of the same length from the same patient. We window the data into 60-second windows with a stride of 30 seconds. Furthermore, we resample all modalities to the highest sampling frequency, in this case 32 Hz, using the SciPy \textit{interp1d} function. We use transformer models for our IIC approach, which were trained in a $10$-fold setting with up to 4 patients per fold. We select the best transformer model from five different initializations for each fold. We report results averaged across all evaluation folds.

\subsection{Proposed Method and Baselines}

For our IIC approach, we ran Algorithm \ref{alg:optimization} for 200 epochs, using the Adam optimizer \cite{kingma2015adam} with a learning rate of $1e^{-2}$. We set the maximum degradation to $0.01$ and set the penalty to $25$.

To the best of our knowledge, there is no direct baseline for our proposed method, as no other XAI method provides local and global concept-based explanations of raw time-series data. Therefore, we compare our method against the closest approaches that yield comparable explanations. We define an approach as a model, combined with a corresponding explanation method. This way, we can assess how adaptations of existing methods would perform in solving the same problem, compared with our proposed method. We do not compare against commonly used, saliency-based methods such as \cite{pmlr-v139-crabbe21a, pmlr-v202-enguehard23a, liu2024explaining}, as they do not yield the global, concept-based explanations we seek for cohort-level interpretation. Furthermore, we do not compare against TCAVs \cite{kim2018interpretability}, as the method does not provide local explanations for patient-specific instance-level interpretation. In our experiments, we use the following baselines.

The first baseline approach is a linear adaptation of CBMs \cite{koh2020concept}, which we refer to as linear CBM (\textbf{LCBM}). While the overall idea of CBMs is to learn concepts from raw time series as a first step, we can skip this step because we have already derived the concepts mathematically in Section \ref{sec:iics}. We will automatically summarize the time-series components to scalar concept values by computing the mean value along the time axis. A vector of those concept values will then be used to classify the input via a linear regression model. The weight vector learned by linear regression can then be used to explain how each component affects the prediction. This makes this baseline inherently interpretable at the group level. 

The second baseline approach combines a fully connected neural network for classification with the widely used SHAP method \cite{lundberg2017unified} for explanations. We refer to this approach as \textbf{FCSHAP}. As input, we provide statistical features derived from the time series to serve as scalar features. In our experiments, we extract the mean, minimum, maximum, and standard deviation. This approach leverages a non-linear, fully connected model while also inheriting all the desirable properties of SHAP explanations. 

\subsection{Quantitative Evaluation}

\textbf{Classification performance} is reported by accuracy and F1-score to evaluate the model's ability to solve the task at hand. To obtain statistically robust performance estimates, we repeated the experiments five times and report the average metrics. For the seizure detection task, we chose different pairings of ictal and interictal segments to assess the generalization over different matches when averaging.

\textbf{Fidelity} assesses the alignment between model behavior and explanations with respect to concepts or features considered important. In our perturbation-based evaluation, we sequentially remove the $k$ most crucial features from the model's input and assess how the model's predictions change by measuring decision flipping \cite{serrano2019attention}. We report the flip rate as the proportion of samples assigned the opposite class label after removing the features. To remove features, we mask them by replacing them with the mean training set time series to maintain physiologically plausible values. 

\textbf{Sufficiency} also assesses the model-explanations alignment, but focusing on the concepts or features deemed unimportant. In a complementary approach to fidelity, we analyze how the model's output changes when we remove all unimportant inputs \cite{deyoung2020eraser}. If the assessment is correct, the network prediction should not change when those inputs are masked out. We again use the flip rate as a sufficiency metric; a lower value indicates better alignment.

\subsection{Qualitative Evaluation}

\textbf{Local explanations} enable the interpretation of explanations from different approaches on exemplary instances. We visualize the recorded wearable data alongside the considered features or components. When comparing importance values, it is important to note that the scales differ due to the normalization constraints imposed by individual XAI methods. Therefore, the focus of the comparison should lie rather in the overall assignment of importance.

\textbf{Global explanations} enable analysis of explanations at the cohort level by indicating how explanations generalize across all patients. We present global explanations of the top $k$ features or components for the dataset in a table. To also analyze the actual values of the components deemed important by IIC, we summarize every time-series component of a sample into a respective scalar by averaging over the time axis. Then, we plot the distribution of the component’s values, filtered to those actually considered by the IIC approach, e.g., those with a score larger than zero.

\section{Results}

\subsection{State Assessment Through Wearables}

\subsubsection{Classification Performance}
Table \ref{tab:performance_eval_state} presents the classification performances for differentiating supine and stepping segments without accelerometer data. This shows that the IIC approach, which uses raw time-series data, performs best, with an average accuracy of $99.0\%$. In comparison, LCBM achieves $81.1\%$ accuracy, while FCSHAP achieves $95.4\%$. The order of approaches is the same when evaluating the F1 scores. IIC has an F1-Score of $99.0\%$, while LCBM and FCSHAP have an F1-Score of $96.1\%$ and $77.7\%$, respectively. The outperformance of IIC over FCSHAP, with a mean difference of $3.6\%$, is statistically significant (p-value = $0.0003$).

\begin{table}[h]
    \centering
    {\small
    \begin{tabular}{|l|c|c|}
    \hline
    \textbf{} & \textbf{Accuarcy} & \textbf{F1} \\
    \hline
    \textbf{IIC} & $\mathbf{99.0}\% \pm \mathbf{0.6}\%$ & $\mathbf{99.0\%} \pm \mathbf{0.6}\%$ \\
    \hline
    \textbf{FCSHAP} & $95.4\% \pm 0.8\%$ & $96.1\% \pm 0.7\%$ \\
    \hline
    \textbf{LCBM} & $81.1\% \pm 2.5\%$ & $77.7\% \pm 0.7$\% \\
    \hline
    \end{tabular}}
    \caption{Evaluation of classification performance for the state assessment task based on wearable autonomic data (no accelerometry data included).}
    \label{tab:performance_eval_state}
\end{table}

\subsubsection{Local Explanations}

\begin{figure}
    \centering
    \includegraphics[width=0.7\linewidth]{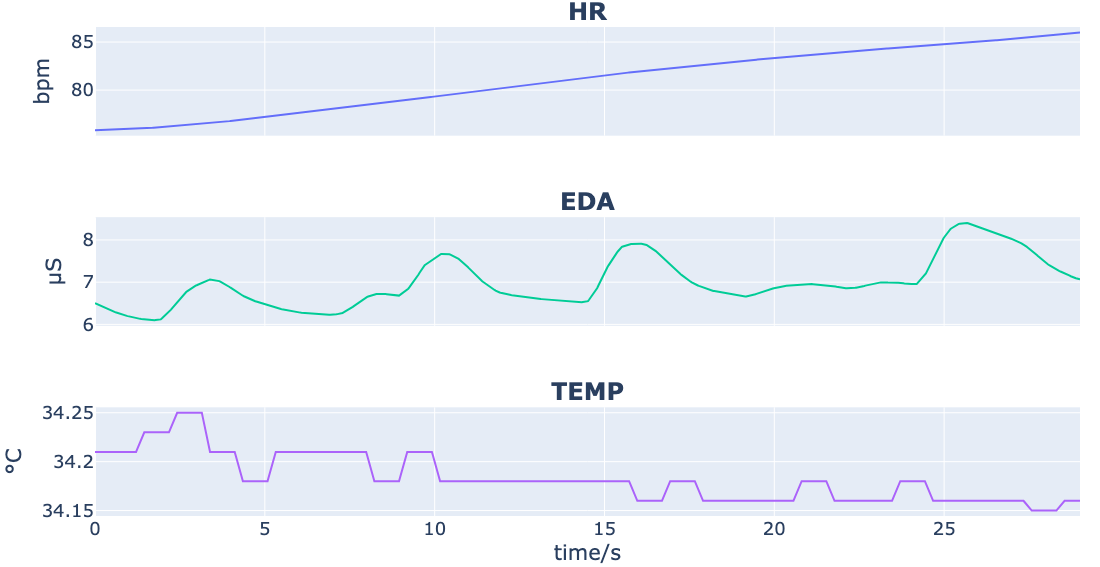}

    \vspace{2mm}

    {\small
    \begin{tabular}{|l|>{\centering\arraybackslash}p{2.cm}|>{\centering\arraybackslash}p{2.71cm}|>{\centering\arraybackslash}p{1.8cm}|}
        \hline
        \textbf{} & \textbf{HR} & \textbf{EDA} & \textbf{TEMP} \\
        \hline
        \textbf{IIC} & Mean oB ($.35$), Variability ($.13$) & Tonic Mean oB ($.69$) & \\
        \hline
        \textbf{LCBM} & Mean oB ($.42$), Variability ($.08$) & Tonic Mean oB ($.20$), Tonic Changes ($.06$) & Rising ($.13$), Falling ($.09$) \\
        \hline
        \textbf{FCSHAP} & Min ($.37$), Mean ($.36$), Max ($.33$) & Max ($.11$), Mean ($.09$), Min ($.06$) & Mean ($.08$), Min ($.06$) \\
        \hline
    \end{tabular}}
    \caption{Exemplary stepping segment that was correctly classified by all approaches, with corresponding explanations (oB denotes over baseline).}
    \label{fig:local_state}
\end{figure}

Figure \ref{fig:local_state} shows an exemplary instance of the stepping class, along with the local concepts and features considered important by the different approaches. All approaches correctly classified this instance. Looking at the time-series example, we can see a heart rate signal that steadily increases from $75$ bpm to $85$ bpm over 30 seconds. The overall trend of the EDA signal is increasing with superimposed smooth tonic fluctuations. It starts at around $6 \mu S$ and reaches more than $8 \mu S$ at the end. The skin temperature signal is relatively stable, fluctuating around $34 ^\circ C$.

When we look at the concepts the IIC approach considers important, the most important component is the tonic mean over baseline of the EDA signal with a score of $0.69$. Subsequently, the mean over baseline and the variability of heart rate are assigned scores of $0.35$ and $0.13$, respectively. IIC does not consider temperature. LCBM considers all three modalities, while heart rate mean over baseline, EDA tonic mean over baseline, and rising temperature are the most important concepts, with scores of $0.42$, $0.2$, and $0.13$, respectively. FCSHAP deems the heart rate signal to be most important, with minimum, mean, and maximum scoring $0.37$, $0.36$, and $0.33$, respectively. EDA tonic maximum is considered as well with $0.11$, while all other features have a score below $0.1$.

\subsubsection{Global Explanations}

\begin{table*}[h]
    \centering
    {\small
    \begin{tabular}{|l|c|c||c|c||c|c|}
    \hline
    \textbf{} & \multicolumn{2}{c||}{\textbf{IIC}} & \multicolumn{2}{c||}{\textbf{LCBM}} & \multicolumn{2}{c|}{\textbf{FCSHAP}} \\
    \hline
    \textbf{k} & \textbf{Component} & \textbf{Import.} & \textbf{Feature} & \textbf{Import.} & \textbf{Feature} & \textbf{Import.}  \\
    \hline
    \textbf{1} & HR Mean oB & $0.516$ & HR Mean oB & $0.377$ & HR Min & $0.302$ \\
    \hline
    \textbf{2} & EDA Tonic Mean oB & $0.137$ & TEMP Rising & $0.126$ & HR Mean & $0.263$ \\
    \hline
    \textbf{3} & TEMP Mean oB & $0.074$ & EDA Tonic Mean oB & $0.118$ & HR Max & $0.215$ \\
    \hline
    \textbf{4} & HR Variability & $0.055$ & TEMP Falling & $0.108$  & TEMP Mean & $0.036$\\
    \hline
    \textbf{5} & TEMP Rising & $0.004$ & EDA Tonic Changes & $0.074$ & EDA Max & $0.029$ \\
    \hline
    \end{tabular}}
    \caption{Top 5 important components and features from the IIC, LCBM, and FCSHAP approaches (oB denotes over baseline).}
    \label{tab:importances_state}
\end{table*}

Table \ref{tab:importances_state} presents the five most important concepts and features across all folds per approach. When comparing the different scores, we can see that the IIC approach based the model output primarily on the mean values of heart rate and EDA, while the temperature mean over baseline and heart rate variability play a minor role. LCBM uses the mean heart rate over baseline as its primary metric and also relies on the rising and falling components of temperature, as well as the tonic mean. Similarly, FCSHAP attribution scores are highest for features related to heart rate levels, namely minimum, mean, and maximum. In addition, the mean temperature and the EDA maximum play a secondary role. 

\begin{figure}[h]
    \centering
    \includegraphics[width=0.7\linewidth]{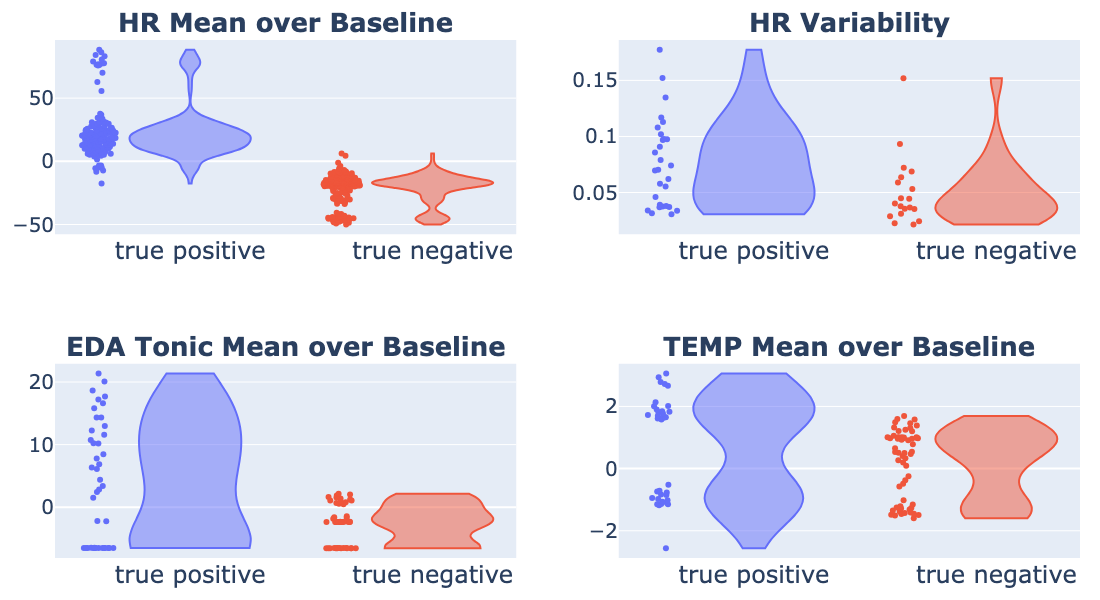}
    \caption{Analysis of the component-wise distribution of scalar summaries, considered from the IIC approach for the state assessment task. We show distributions for true positives (stepping) and true negatives (supine), separately.}
    \label{fig:global_state_values}
\end{figure}
Figure \ref{fig:global_state_values} shows the component-wise distribution of scalar summaries, divided into true-positive (stepping) and true-negative (supine) cases. We can see that the heart rate mean over baseline, when considered by IIC, was higher for stepping segments and lower for resting segments. We can observe similar trends in heart rate variability and EDA tonic mean relative to baseline, with the distributions giving greater weight to higher values for true positives. In contrast, the distribution of the temperature mean relative to baseline values is rather bimodal, with high density around high and low values.

\subsubsection{Numerical Evaluation of Explanations}
Table \ref{tab:fidelity_sufficiency_state} presents the fidelity and sufficiency by flip rate. We can see that the IIC and LCBM approaches have a very high flip rate from $k=1$, indicating that removing the most important concept results in classification performance at the chance level. For FCSHAP, we see an increasing flip rate, starting at $27.9\%$, rising to $29.5\%$, and reaching $55.5\%$, indicating chance-level classification performance at $k=3$. To also assess the complementary property, we computed sufficiency by measuring output degradation when removing features or concepts deemed unimportant. We chose importance thresholds adapted to the approaches' importance-distribution, with $0.01$ for IIC, $0.01$ for LCBM, and $0.02$ for FCSHAP. We can see that the flip rate for both concept-based approaches is very low, at $0.2\%$ and $0.7\%$ for IIC and LCBM, respectively, while FCSHAP's flip rate is slightly higher at $3.8\%$.

\begin{table}[h]
    \centering
    {\small
    \begin{tabular}{|l|c|c|c|}
    \hline
    \multicolumn{4}{|c|}{\textbf{Fidelity}} \\
    \hline
    \textbf{k} & \textbf{IIC} & \textbf{LCBM} & \textbf{FCSHAP} \\
    \hline
    \textbf{1} & $51.2\%$ & $49.5\%$ & $27.9\%$ \\
    \hline
    \textbf{2} & $52.0\%$ & $49.5\%$ & $29.5\%$ \\
    \hline
    \textbf{3} & $50.2\%$ & $49.5\%$ & $55.5\%$  \\
    \hline
    \hline
    \multicolumn{4}{|c|}{\textbf{Sufficiency}} \\
    \hline
    & $0.2\%$ & $0.7\%$ & $3.8\%$ \\
    \hline
    \end{tabular}}
    \caption{Evaluation of fidelity and sufficiency by measuring the flip rate for the state assessment task.}
    \label{tab:fidelity_sufficiency_state}
\end{table}

\subsection{Epileptic Seizure Detection with Wearables}

\subsubsection{Classification Performance}
The results in Table \ref{tab:performance_eval_sz} show accuracy and F1-score averaged over different ictal-interictal matchings. The IIC approach using raw time-series data achieves an average accuracy of $87.8\%$, which is higher than that of the FCSHAP approach ($82.6\%$). The LCBM approach yields an average accuracy of $63.6\%$. The outperformance of IIC over FCSHAP with an average improvement of $5.2\%$ is statistically significant (p-value = $0.0005$). Due to the LCBM approach's comparatively low performance, we will omit its explanations and analysis in the following, as we do not expect its model to have learned something worth interpreting.

\begin{table}[h]
    \centering
    {\small
    \begin{tabular}{|l|c|c|}
    \hline
    \textbf{} & \textbf{Accuracy} & \textbf{F1} \\
    \hline
    \textbf{IIC} & $\mathbf{87.8\%} \pm \mathbf{1.4}\% $ & $\mathbf{86.7}\% \pm \mathbf{2.1}\%$ \\
    \hline
    \textbf{FCSHAP} & $82.6\% \pm 2.7\%$ & $81.6\% \pm 1.9\%$ \\
    \hline
    \textbf{LCBM} & $63.6\% \pm 2.5\%$ & $63.7\% \pm 2.7\%$\\
    \hline
    \end{tabular}}
    \caption{Comparison of classification performance for the seizure detection task.}
    \label{tab:performance_eval_sz}
\end{table}

\subsubsection{Local Explanations}

\begin{figure}
    \centering
    \includegraphics[width=0.7\linewidth]{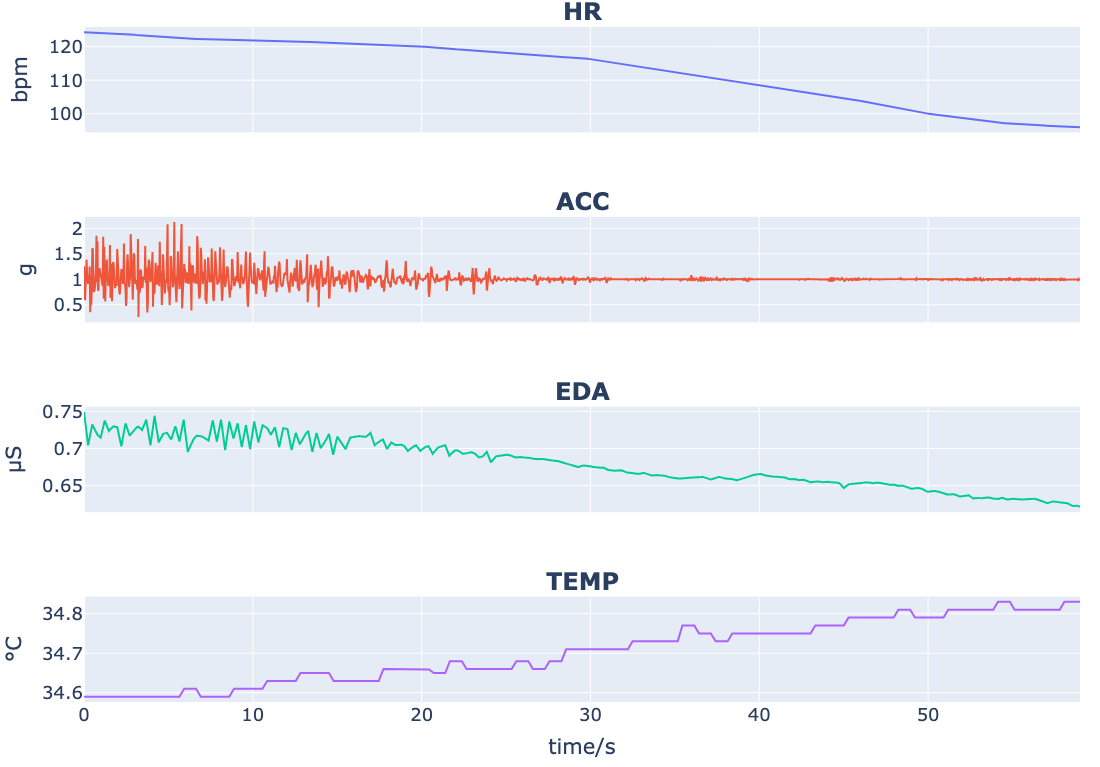}

    \vspace{2mm}

    {\small
    \begin{tabular}{|l|>{\centering\arraybackslash}p{3cm}|>{\centering\arraybackslash}p{3.9cm}|}
        \hline
        \textbf{} & \textbf{HR} & \textbf{ACC} \\
        \hline
        \textbf{IIC} & Mean oB ($.97$), Variability ($.11$) & Outlier ($.98$), Activity ($.76$) \\
        \hline
        \textbf{FCSHAP} & Max ($.14$), Mean ($.12$), Min ($.05$), Std ($.03$) & Std ($.27$), Min ($.07$), Max ($.03$) \\
        \hline
    \end{tabular}}
    \caption{Exemplary seizure segment that was correctly classified as a seizure by both approaches, with corresponding explanations (oB denotes over baseline).}
    \label{fig:seiz_local_tp}
\end{figure}

Figure \ref{fig:seiz_local_tp} shows a visualization of wearable data recorded during a seizure with corresponding component and feature importances. The heart rate signal exhibits a comparatively high average and descents from over 120 bpm to near 100 bpm. The accelerometry signal shows substantial variance in the first half of the window, with values ranging from $0.3g$ to $2g$, and the variance decreases over time. Next, the EDA signal is decreasing from $0.75 \mu S$ to around $0.6 \mu S$ at the end of the window. Furthermore, it contains noise at the beginning, whereas the accelerometry signal exhibits variance. Lastly, the temperature signal shows a slight increase in skin temperature of $0.2^\circ C$. Both approaches correctly classified this seizure example. The IIC approach indicates that only the heart rate and accelerometry modalities have nonzero weights. For heart rate, the mean relative to the baseline component was by far the most important, followed by the variability component. For accelerometry, the outlier and the activity component were both considered important, whereas the mean component was not. The FCSHAP approach assigns equal importance to the same modalities: maximum, mean, minimum, and standard deviation for heart rate, and scores for standard deviation, minimum, and maximum for the accelerometry. Either approach considered neither EDA nor temperature-related components or features.

\begin{figure}
    \centering
    \includegraphics[width=0.7\linewidth]{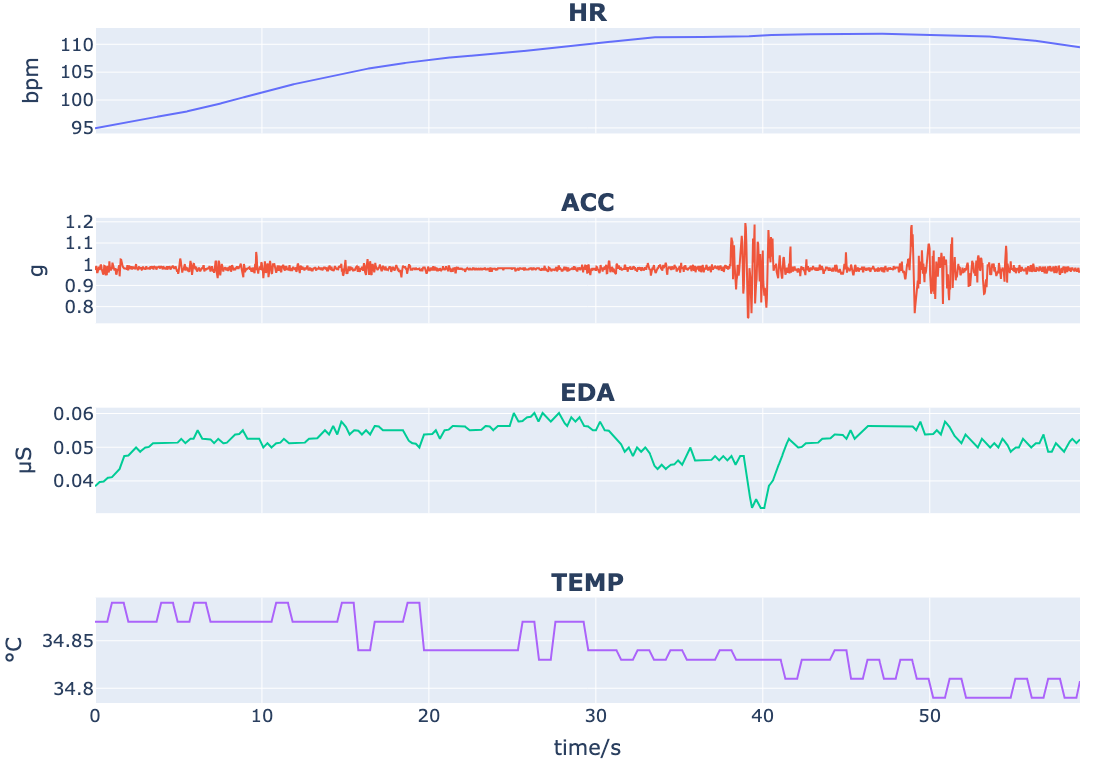}

    \vspace{2mm}
    
    {\small
    \begin{tabular}{|l|>{\centering\arraybackslash}p{3cm}|>{\centering\arraybackslash}p{3.9cm}|}
        \hline
        \textbf{} & \textbf{HR} & \textbf{ACC} \\
        \hline
        \textbf{IIC} & Mean oB ($.79$), Variability ($.36$) & Mean oB ($.56$), Outlier ($.32$) \\
        \hline
        \textbf{FCSHAP} & Min ($.18$), Max ($.15$), Std ($.14$), Mean ($.14$) & Max ($.2$), Std ($.16$), Min ($.04$) \\
        \hline
    \end{tabular}}
    \caption{Exemplary interictal segment that was classified as a seizure segment by both approaches, with corresponding explanations (oB denotes over baseline).}
    \label{fig:seiz_local_fp}
\end{figure}

Figure \ref{fig:seiz_local_fp} illustrates a false positive: an interictal segment classified as a seizure. The heart rate signal shows a rising trend, starting at $95$ bpm and peaking at $112$ bpm. The accelerometry signal shows two regions of high variance in the second half of the window, with values ranging from $0.7g$ to $1.2g$, whereas elsewhere the variance is lower. The EDA signal changes around a mean of $0.05 \mu S$, with small dips and rises distributed throughout the window. Lastly, the temperature signal shows a relatively constant skin temperature of $34.8 ^\circ C$. The corresponding explanations from both approaches are added. Results from the IIC approach score the mean and variability components of heart rate, as well as gravitation and outlier components from accelerometry. The FCSHAP approach assigns importance to the mean, minimum, maximum, and standard deviation of heart rate. Furthermore, the accelerometer's maximum, standard deviation, and minimum are assigned appropriate attributes. As with the true positive example, EDA and temperature modalities were not considered important in this decision.

\subsubsection{Global Explanations}

Table \ref{tab:importances_seizure} presents the 10 most important components and features identified by the IIC and FCSHAP approaches. The IIC approach assigns the highest weights to the accelerometry outlier, the mean heart rate relative to baseline, the mean EDA Tonic relative to baseline, and the heart rate variability components. In contrast, the FCSHAP approach attributes most to the standard deviation of accelerometry, as well as to the maximum, minimum, and mean heart rates.

\begin{table}[h]
    \centering
    {\small
    \begin{tabular}{|l|c|c||c|c|}
    \hline
    \textbf{} & \multicolumn{2}{c||}{\textbf{IIC}} & \multicolumn{2}{c|}{\textbf{FCSHAP}} \\
    \hline
    \textbf{k} & \textbf{Component} & \textbf{Import.} & \textbf{Feature} & \textbf{Import.} \\
    \hline
    \textbf{1} & ACC Outlier & $0.421$ & ACC Std & $0.233$ \\
    \hline
    \textbf{2} & HR Mean oB & $0.342$ & HR Max & $0.176$ \\
    \hline
    \textbf{3} & EDA Tonic Mean oB & $0.262$ & HR Min & $0.101$ \\
    \hline
    \textbf{4} & HR Variability & $0.220$ & HR Mean & $0.098$ \\
    \hline
    \textbf{5} & ACC Activity & $0.214$ & HR Std & $0.065$ \\
    \hline
    \textbf{6} & ACC Mean oB & $0.135$ & EDA Max & $0.053$ \\
    \hline
    \textbf{7} & TEMP Mean oB & $0.127$ & ACC Max & $0.050$ \\
    \hline
    \textbf{8} & EDA Tonic Ch. & $0.019$ & ACC Min & $0.044$ \\
    \hline
    \textbf{9} & TEMP Rising & $0.012$ & EDA Min & $0.023$ \\
    \hline
    \textbf{10} & TEMP Falling & $0.009$ & TEMP Max & $0.022$ \\
    \hline
    \end{tabular}}
    \caption{Top 10 components and features from the IIC and FCSHAP approaches in decreasing order (oB denotes over baseline).}
    \label{tab:importances_seizure}
\end{table}

\begin{figure}[h]
    \centering
    \includegraphics[width=0.7\linewidth]{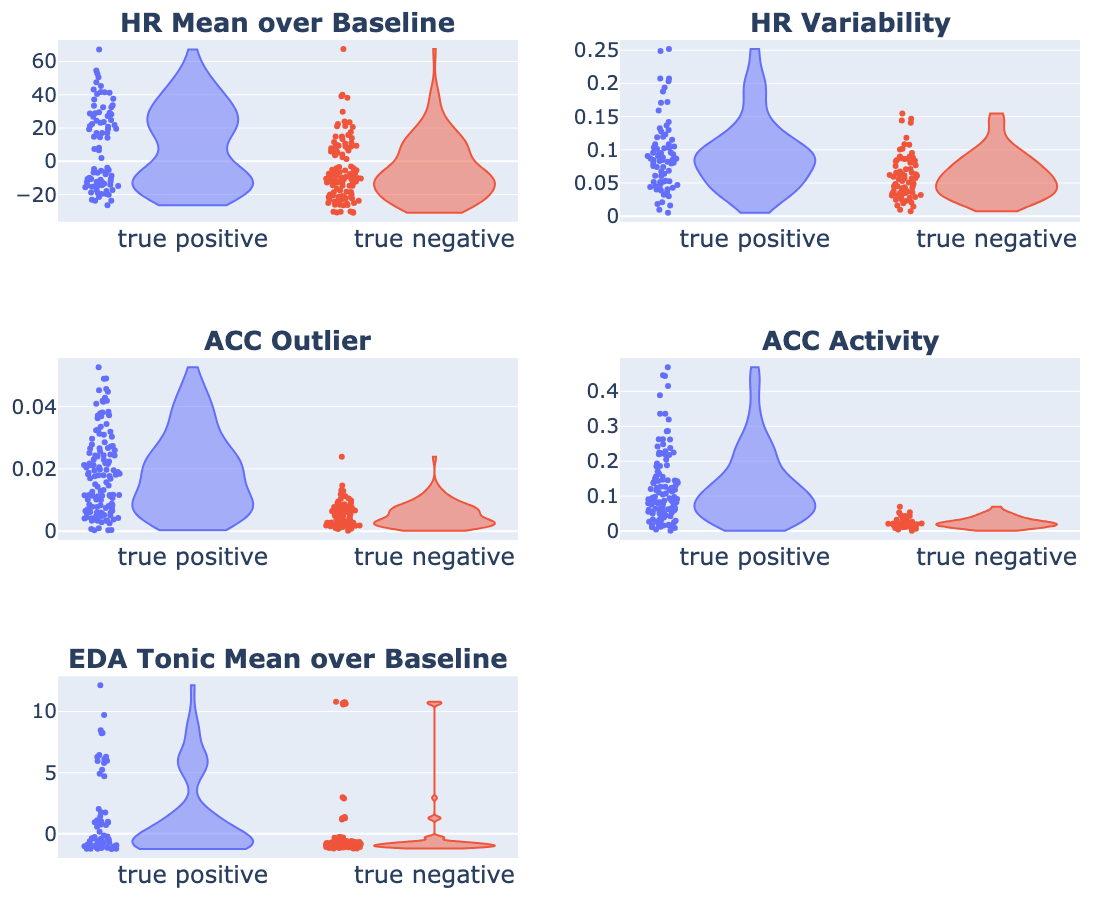}
    \caption{Analysis of the component-wise distribution of scalar summaries, considered from the IIC approach for the seizure detection task. We show distributions for true positives (seizure) and true negatives (no seizure), separately.}
    \label{fig:global_sz_values}
\end{figure}

To conclude the global analysis, Figure \ref{fig:global_sz_values} shows the distribution of scalar summaries across all components identified as important by IIC. We observe that the considered outlier and activity values are higher for true positives than for true negatives. The same holds for the tonic mean values from the EDA, relative to the baseline. In heart rate, the mean and variability components show greater overlap between true positives and negatives, but the density of higher values is higher for true positives.

\subsubsection{Numerical Evaluation of Explanations}
Table \ref{tab:fidelity_sufficiency_seizure} compares the fidelity of both approaches when omitting the first $k$ components or features, cumulatively. We observe that the flip rate for the IIC approach increases smoothly with the number of components omitted, from $9.7\%$ to $38.2\%$ as $ k$ increases from $1$ to $6$. For the FCSHAP approach, the drop is higher for the first feature, starting at $36.1\%$, but the increase with the number of features left out is smaller, reaching $46.7\%$ at $k=6$. The sufficiency results report the flip rate when masking components and features deemed unimportant. We used thresholds to define sets of unimportant features, adapted to the full distribution of importances. The results show that the IIC approach's predictions do not flip at all, with an average flip rate of $0\%$. The FCSHAP approach's flip rate is $6.3\%$.

\begin{table}[h]
    \centering
    {\small
    \begin{tabular}{|l|c|c|}
    \hline
    \multicolumn{3}{|c|}{\textbf{Fidelity}} \\
    \hline
    \textbf{k} & \textbf{IIC} & \textbf{FCSHAP} \\
    \hline
    \textbf{1} & $9.7\%$ & $36.1\%$ \\
    \hline
    \textbf{2} & $16.4\%$ & $41.0\%$ \\
    \hline
    \textbf{3} & $15.5\%$ & $42.0\%$ \\
    \hline
    \textbf{4} & $19.9\%$ & $40.0\%$ \\
    \hline
    \textbf{5} & $33.9\%$ & $42.0\%$ \\
    \hline
    \textbf{6} & $38.2\%$ & $46.7\%$ \\
    \hline
    \hline
    \multicolumn{3}{|c|}{\textbf{Sufficiency}} \\
    \hline
    & $0\%$ & $6.3\%$ \\
    \hline
    \end{tabular}}
    
    \caption{Evaluation of fidelity and sufficiency by measuring the flip rate for the seizure detection task.}
    \label{tab:fidelity_sufficiency_seizure}
\end{table}

\section{Discussion}

We present an XAI method that combines the desirable properties of saliency-based and concept-based methods: it leverages raw time-series inputs to achieve optimal performance and provides concept-based explanations for interpretability. For both datasets, we observe outperformance over using extracted features with fully connected neural networks, while maintaining interpretability. In our experiments, we used IIC in combination with LSTM and transformer architectures, demonstrating its model-agnostic advantage.

\subsection{Conceptual Comparison}

A conceptual comparison of the different approaches and their properties is presented in Table \ref{tab:conceptual_comparison}. As the approaches consider not only different XAI methods but also different models, our goal is to examine the advantages and limitations of these approaches, including the trade-off between model performance and explanation quality. The LCBM approach is the only one of the three where the underlying model is interpretable by default. Using scalar concept values in linear regression yields importance scores from the regression coefficients. As a downside, this approach is prone to underperforming native time-series neural networks, as it is restricted to linear relations and relies on only a set of scalar concepts. Therefore, the LCBM approach's linearity restriction imposes a trade-off between performance and interpretability, favoring interpretability. The FCSHAP approach overcomes these performance limitations by utilizing non-linear relationships via a fully connected neural network. This comes at the cost of losing interpretability on the model side. SHAP can be used to generate explanations, enabling the interpretation of black-box models. The drawback, however, is that this approach cannot fully exploit all the information in the raw time-series input, as a set of extracted features must be fed to the network. That is why this approach is prone to underperforming an LSTM or transformer model, which leverages all information in the raw data. The proposed IIC approach enables the use of time-series-native models, such as RNNs and transformers, without compromising performance. While those models are also black boxes, explanations via IICs enable domain experts to interpret them using concepts of interest. Furthermore, the IIC approach allows for a post-hoc interpretability, in which the components can still be changed and refined. In contrast, both LCBM and FCSHAP need the concepts and features during training. One drawback, however, is the restriction on components, as the IIC approach requires an invertible transformation from the time domain to the component domain. LCBM and FCSHAP both have no requirements for the features used, enabling the use of non-invertible transformations.

\begin{table}[h]
    \centering
    {\small
    \begin{tabular}{|l|c|c|c|}
    \hline
    \textbf{Properties} & \textbf{Proposed IIC}& \textbf{LCBM} & \textbf{FCSHAP} \\
    \hline
    \textbf{Time-Series Native} & \checkmark & \texttimes & \texttimes \\
    \hline
    \textbf{Interpretable model} & \texttimes & \checkmark & \texttimes \\
    \hline
    \textbf{Local Explanations} & \checkmark & \texttimes & \checkmark \\
    \hline
    \textbf{Global Explanations} & \checkmark & \checkmark & \checkmark \\
    \hline
    \textbf{Label-independent} & \checkmark & \texttimes & \texttimes \\
    \hline
    \textbf{Model-independent} & \checkmark & \texttimes & \texttimes \\
    \hline
    \textbf{Arbitrary Features} & \texttimes & \checkmark & \checkmark \\
    \hline
    \textbf{Post-hoc Adaptable} & \checkmark & \texttimes & \texttimes \\
    \hline
    \end{tabular}}
    \caption{Comparison of the approaches and their properties.}
    \label{tab:conceptual_comparison}
\end{table}

\subsection{Classification Performance}
For both datasets, we evaluated the classification performances of the different approaches. On the state assessment task, the LSTM model achieved $99.0\%$ accuracy, significantly outperforming both LCBM ($81.1\%$) and FCSHAP ($95.4\%$) by $17.9\%$ and $3.6\%$, respectively. Although the feature-based FCSHAP approach demonstrated competitive performance, the significant outperformance indicates that relying on features still entails a performance trade-off. For the second dataset, we employed a transformer model to detect seizures in raw time-series data from clinical recordings. We found that the transformer outperformed both competitors significantly, achieving an average accuracy of $87.4\%$ compared to $83.4\%$ and $63.6\%$ for FCSHAP and LCBM, respectively, across multiple ictal-interictal instance pairings. This reinforces our hypothesis that using raw time-series data offers meaningful advantages over feature-extraction approaches. 

While transitioning from time series to features may improve interpretability, our findings highlight a critical trade-off in performance. This trade-off carries particular weight in medical domains: although interpretability remains essential, the cost of reduced accuracy directly impacts patient safety. In seizure detection specifically, each correct alarm not only reduces the likelihood of seizure-related injuries but also increases the probability that caregivers can provide timely intervention and support \cite{hadady2023real}. Since prompt caregiver response can be crucial in preventing serious complications - including sudden unexpected death in epilepsy (SUDEP) - every percentage point of accuracy might have meaningful implications for patient health and well-being.

\subsection{Local and Global Explanations}
We interpreted local, instance-wise, and global, cohort-level explanations provided for both datasets and assessed them against existing knowledge in the domains. For the state assessment task, we found that all approaches used similar reasoning in their models, with heart rate as the primary factor distinguishing between resting in the supine position and stepping. All approaches found that the overall heart rate level was most helpful. The IIC and LCBM approaches captured that information via the heart rate mean over baseline component, while FCSHAP used the heart rate mean, minimum, and maximum. Given the task at hand, relying on these patterns is justified, as cardiac activity is higher during active stepping than during supine rest \cite{Wilson2025Abnormal}. Further, all approaches consider the absolute level of the EDA. While IIC and LCBM use the EDA tonic mean over baseline, FCSHAP uses the EDA maximum, thereby capturing similar information. Temperature is considered as well: IIC uses the mean temperature over baseline, FCSHAP uses the mean temperature, and LCBM uses the rising and falling components. Both EDA and temperature are related to physical activity \cite{Wilson2025Abnormal} and therefore provide a reasonable indication of physical activity. This result supports our hypothesis that inherently interpretable components can yield concept-based explanations at both local and global levels. At the same time, the model still uses raw time sequences and outperforms competitors in classification.

For the seizure-detection task, local and global explanations indicate that both approaches capture the relevant information in the data to identify a seizure. As the data include only tonic-clonic seizures, we expect a focus on accelerometry that captures the repetitive motor movements characteristic of seizures \cite{kodankandath2023generalized}. A steep increase in heart rate is also reported in the literature \cite{devinsky2004effects}, which is why we expect the absolute level and variability to capture relevant information for classification. This is exactly what we found when examining local and global explanations for correctly classified seizures. Both approaches captured those markers in their respective ways. The summary of actual components in Figure \ref{fig:global_sz_values} further confirms this by showing increased values in the respective components considered by the transformer. There is also a known relationship between tonic-clonic seizures and increased sweating \cite{devinsky2004effects}; however, due to the latency of sweating, we do not expect it to be present in every seizure segment. This is reflected by the explanations of both approaches, with EDA being relevant. The summary of component values verifies this by showing a heavier tail for large values than for all samples. The tonic change and phasic components, however, do not appear to be relevant for detection. Regarding the skin temperature, there is no established biomarker in the literature for seizure detection with wearables. Therefore, we do not expect it to be relevant to the given task. However, we can see that the modality is not irrelevant. This could point towards a cofounder in the data, especially when we randomly pair ictal and interictal segments. Further effects could arise from the recording setup in the long-term monitoring unit, where patients may have their wrists covered with a blanket, potentially affecting skin temperature measurements.

Our findings support the hypothesis that, even though the LSTM and the transformer model were trained on raw time-series data, we can interpret them using domain-specific concepts of interest via IICs. Compared with the baseline explanations from the FCSHAP and LCBM approaches, our method provides interpretability comparable to theirs, with importance scores encoding equivalent information from the raw time-series data. We showed a cohort-level interpretation of the models, yielding an overall understanding of the learned connections. Moreover, through local explanations, our method enables the interpretation of individual predictions of interest, such as false positives — a capability of particular clinical importance for seizure detection, where understanding erroneous detections is vital for establishing clinician trust and ensuring appropriate patient care.

\subsection{Numerical Evaluation of Explanations}
We evaluated the fidelity and sufficiency of the different explanations for both datasets. For the state assessment task, the IIC and LCBM approaches already had very high flip rates at $k=1$, leading to chance-level performance without the most important concept. The FCSHAP approach also reached chance-level for $k=3$. Further, the flip rates in the sufficiency results were very small. For the seizure detection task, we observe a gradual increase in fidelity for the IIC approach, measured by flip rate from $9.7\%$ at $k=1$ to $38.2\%$ at $k=6$. The increase is smoother than that of Feature-SHAP, starting at $36.1\%$, suggesting that the transformer model is more robust and has learned a different interplay of components and modalities, relying on multiple components rather than a single one. The sufficiency remains low.

The numerical evaluation shows that the IIC explanation approach yields explanations that align well with the actual model behavior. When omitting important features one by one, the model's performance drops considerably. Components deemed unimportant do not reduce performance when left out. This validates that the IIC approach's explanations can be used to interpret state-of-the-art architectures, such as transformers, without sacrificing performance. 

\subsection{Selecting Components}

The choice of IICs is essential for the explanations to be intuitive. The proposed decomposition captures multiple components discussed and used in the literature on health monitoring with wearables. However, the exact decompositions should be application-specific to capture the relevant patterns. To do so, the proposed components can be further decomposed, e.g., by separating ACC activity into rhythmic and arrhythmic components to better capture shaking in motor-based seizures. Furthermore, frequency information is commonly considered in decompositions, as in \cite{brusch2024flextime}. As the transformation to the frequency domain is invertible, frequency components can also be utilized with our method. However, frequency-domain features from wearable data are not inherently interpretable, as interpretation depends heavily on the sensor. Selection and cut-off values may require input from healthcare or sports professionals, along with understanding related datasets, to improve interpretability. Decompositions, in turn, may provide clinicians, healthcare professionals, and patients with actionable insight into the most prominent clinical variables, ultimately enabling more personalized monitoring and modeling of clinical variables towards individualized and precision medicine. In future work, we would like to explore IICs in combination with EEG frequency bands. It is important to ensure that the choice of components does not lead to confirmation bias. Simply validating the expected components might lead to circular logic and cause us to overlook the most insightful novel or counterintuitive components. Therefore, we recommend applying related components that may otherwise be considered irrelevant.

In wearable data, quality is essential to all algorithm development. As shown in \cite{bottcher2022data}, data quality impacts both overall performance and interpretability. While poor-quality data remains an inherent limitation, our method is robust to noise because all decompositions are explicitly defined, ensuring the resulting components are reliable with respect to their definition.

\section{Conclusion}
\label{sec:conclusion}

We propose a novel, model-agnostic, concept-based XAI method designed for models trained on raw time-series data. By IICs, we can generate local, window-level explanations as well as global, cohort-level explanations, without a performance-interpretability tradeoff. Furthermore, we propose decomposing multimodal wearable data into IICs, custom-tailored for health-monitoring applications. We evaluate the combination of a time-series model (in our case, an LSTM and a transformer) and explanations via IICs across two clinical health-monitoring datasets. We compare our proposed technique against the closest approaches that yield concept- or feature-based explanations. Across datasets, our approach outperforms competitors in classification accuracy, achieving $99.0\%$ on the state assessment task and $87.8\%$ on the seizure detection task, thereby demonstrating the superiority of time-series-based models for these tasks. At the same time, our IIC approach yields accessible, relevant explanations for the tasks at hand, which can help domain experts interpret the trained model at multiple levels. We further validated that the explanations align with the model's behavior by numerically assessing their fidelity and sufficiency. In future work, we aim to evaluate the proposed method for explaining representations in unsupervised settings \cite{kuschel2025explaining} and to develop components for other types of time series, such as EEG and ECG.

\section{Acknowledgements}

The study was supported by the Paderborn University Research Award and partially by the Epilepsy Research Fund and by the Patricia B. Terwilliger Family Chair in Epilepsy Research at Boston Children's Hospital. SV was supported by the High-tech Agenda Bavaria. We thank the team at the Sports Medicine Group at Paderborn University for collecting the state assessment data, and the team at the Division of Epilepsy and Clinical Neurophysiology at Boston Children's Hospital for collecting the seizure data.

\section{Declaration of Competing Interest}

The authors declare the following financial interests/personal relationships, which may be considered potential competing interests: Tobias Loddenkemper is a co-inventor on pending and approved patents for detecting and predicting clinical outcomes, as well as for detecting, managing, diagnosing, and treating neurological conditions, including epilepsy and seizures. He received device donations for research purposes from various companies, including Empatica. In the past, he received research support from Empatica, paid to Boston Children’s Hospital, for research unrelated to this study. Tanuj Hasija, Maurice Kuschel, Claus Reinsberger, Tobias Loddenkemper and Solveig Vieluf are part of pending patent applications covering technology for seizure forecasting and explainable seizure monitoring.

\bibliographystyle{unsrt}  
\bibliography{final_references} 

\end{document}